\documentclass[preprint,review,12pt]{elsarticle}
\usepackage{lineno}
\linenumbers
\usepackage{amsmath}	

\usepackage{amssymb}

\journal{Nuclear Physics B}

\begin{document}

\begin{frontmatter}



\title{Proton Radiation Damage Experiment on P-Channel CCD for an X-ray CCD
 camera onboard the Astro-H satellite} 


\author[label1]{Koji Mori}
\author[label1,label2]{Yusuke Nishioka}
\author[label1]{Satoshi Ohura}
\author[label1]{Yoshiaki Koura}
\author[label1]{Makoto Yamauchi}

\author[label3]{Hiroshi Nakajima}
\author[label3]{Shutaro Ueda}
\author[label3]{Hiroaki Kan}
\author[label3]{Naohisa Anabuki}
\author[label3]{Ryo Nagino}
\author[label3]{Kiyoshi Hayashida}
\author[label3]{Hiroshi Tsunemi}

\author[label4]{Takayoshi Kohmura}
\author[label4]{Shoma Ikeda}

\author[label5]{Hiroshi Murakami}

\author[label6]{Masanobu Ozaki}
\author[label6]{Tadayasu Dotani}

\author[label7]{Yukie Maeda}
\author[label8]{Kenshi Sagara}

\address[label1]{Department of Applied Physics and Electronic Engineering,
 Faculty of Engineering, University of Miyazaki, 1-1 Gakuen Kibanadai-Nishi,
 Miyazaki, 889-2192, Japan} 

\address[label2]{Technical center,
 Faculty of Engineering, University of Miyazaki, 1-1 Gakuen Kibanadai-Nishi,
 Miyazaki, 889-2192, Japan} 

\address[label3]{Department of Earth and Space Science, Graduate School of Science, Osaka
 University, 1-1 Machikaneyama, Toyonaka, Osaka, 560-0043, Japan}

\address[label4]{Physics Department, Kogakuin University, 2665-1, Nakano,
 Hachioji, 192-0015, Japan} 

\address[label5]{Department of Physics, Faculty of Science, Rikkyo University,
 3-34-1, Nishi-Ikebukuro, Toshima-ku, Tokyo, 171-8501, Japan} 

\address[label6]{Institute of Space and Astronautical Science, Japan Aerospace
 Exploration Agency, 3-1-1 Yoshinodai, Chuo-ku, Sagamihara, Kanagawa
 252-5210, Japan} 

\address[label7]{Faculty of Engineering, University of Miyazaki, 1-1 Gakuen
 Kibanadai-Nishi, Miyazaki, 889-2192, Japan}

\address[label8]{Department of Physics, Kyushu University, 6-10-1 Hakozaki,
 Higashi-ku, Fukuoka 812-8581, Japan}  

\begin{abstract}
We report on a proton radiation damage experiment on P-channel CCD newly
developed for an X-ray CCD camera onboard the Astro-H satellite. The
device was exposed up to 10$^9$ protons~cm$^{-2}$ at 6.7~MeV. The
charge transfer inefficiency (CTI) was measured as a function of
radiation dose. In comparison with the CTI currently measured in the CCD
camera onboard the Suzaku satellite for 6~years, we confirmed that the new
type of P-channel CCD is radiation tolerant enough for space use. We
also confirmed that a charge-injection technique and lowering the operating
temperature efficiently work to reduce the CTI for our device. A
comparison with other P-channel CCD experiments is also discussed.
\end{abstract}


\end{frontmatter}


\section{Introduction}
\label{intro}

Charge-coupled devices (CCDs) have an almost 20-years long history as space-borne
detectors for X-ray astronomy. The ASCA satellite for the first time employed X-ray
photon counting CCDs\cite{1994PASJ...46L..37T}, which were front-illuminated (FI)
devices with a depletion layer thickness of about 30~$\mu$m. Subsequent Japanese
X-ray satellite Suzaku carries, in addition to FI CCDs, a back-illuminated (BI) CCD
that significantly improved the detection efficiency for soft X-ray photons down to
0.3~keV\cite{2007PASJ...59S..23K}. However, the depletion layer thickness of the BI
CCD was still limited to about 40~$\mu$m. The X-ray CCDs flown to space so far were
all made from P-type silicon wafers, namely N-channel CCDs. Recently, a new type of
P-channel CCD has become available with a thick depletion layer of a few hundred
$\mu$m\cite{2006JaJAP..45.8904M}. We employ the new P-channel CCD for Soft X-ray
Imager (SXI)\cite{2010SPIE.7732E..28T,2011SPIE.8145E.239H}, a new CCD camera onboard
the upcoming Astro-H satellite\cite{2010SPIE.7732E..27T}. Using the P-channel CCD as
a BI device with a depletion layer thickness of 200~$\mu$m, high detection
efficiency for both hard and soft X-ray photons can be achieved.

Since P-channel CCDs have no performance experience in space, their radiation
hardness is an issue to be examined before launch. The primary source of radiation
damage of CCD is cosmic-ray protons, which produce displacement damage in silicon
resulting in the formation of carrier traps. Traps in the channel region capture
charge carriers during transfer. Then, the charge transfer inefficiency (CTI), a
fraction of charge loss per one pixel transfer, is frequently used as a measure of
radiation damage for X-ray CCDs.  There are several experimental reports indicating
that P-channel CCDs are actually radiation harder than conventional N-channel CCDs
in terms of the
CTI\cite{1999ITNS...46..1790N,2002ITNS...49..1221N,2004SPIE.5499..542M}. The greater
radiation hardness of P-channel CCDs may be explained by the difference of carrier
traps in the two different type of
CCDs\cite{1997spratt,1999ITNS...46..1790N}. Proton-induced formation of divacancy
hole traps is considered to be less favorable in a P-channel CCD compared to that of
phosphorus-vacancy electron traps in an N-channel
CCD\cite{2002ITNS...49..1221N}. However, it is also suggested that other kind of
traps are possibly produced and they may adversely increase the CTI in a P-channel
CCD\cite{2001TNS...48..960N}.

In relation to the radiation hardness of the device, mitigating the
radiation damage effect is also important. A charge-injection (CI)
technique is one of such mitigation methods\cite{2007SPIE.6686E..23B,
2008PASJ...60S...1N, 2009PASJ...61S...9U}. In this technique, charges
are intentionally injected to selected rows which are regularly
spaced. Then, the injected charges work as sacrifices to fill traps and
following real X-ray-induced charges are transferred with lesser charge
loss. The CI has been verified to effectively reduce the CTI in
the case of the Suzaku CCDs\cite{2007SPIE.6686E..23B}. Lowering operating
temperature of CCDs also reduced the CTI in the case of the Chandra and
XMM-Newton CCDs\cite{2004SPIE.5488..264S,2006SPIE.6276E..48G}.

In this paper, we report on a proton radiation damage experiment on our
newly developed P-channel CCD. We here describe how radiation-hard our
new device is and how our mitigation methods work once the device is
damaged. 

\section{Experiment}

\subsection{P-channel CCD used in this experiment} 

We developed a P-channel BI CCD for the SXI, Pch-NeXT4, in collaboration with
Hamamatsu Photonics K.K.\cite{2010SPIE.7732E..28T,2011SPIE.8145E.239H} (``NeXT'' is
the former project name of the Astro-H satellite). The Pch-NeXT4 is a frame transfer
type CCD with an imaging area of 30.72~mm square. There are 1280$\times$1280
physical pixels in the imaging area, which will be 640$\times$640 logical pixels
after on-chip 2$\times$2 binning. There are four readout nodes, and we nominally use
two of them.

There were two CCDs used in this experiment. One was the same type of CCD as
Pch-NeXT4, and the other was a smaller size CCD but made in the same
manufacturing process as that of Pch-NeXT4. We hereafter refer to the former and
the latter as large-CCD and mini-CCD, respectively. 

\subsection{Experimental setup} 

We performed our proton radiation damage experiment at the Kyushu
University tandem accelerator laboratory. The proton beam current was
50~nA--1~$\mu$A. Since the direct beam intensity was too strong for our
purpose even at a low current of 10~nA (10~year in-orbit equivalent
protons would be irradiated in less than one second), we used scattered
protons. Use of scattered protons provided another benefit that the
large-CCD could be almost uniformly damaged as is the actual case in
space.

\begin{figure}[ht]
 \centering
 \includegraphics[width=1.0\linewidth]{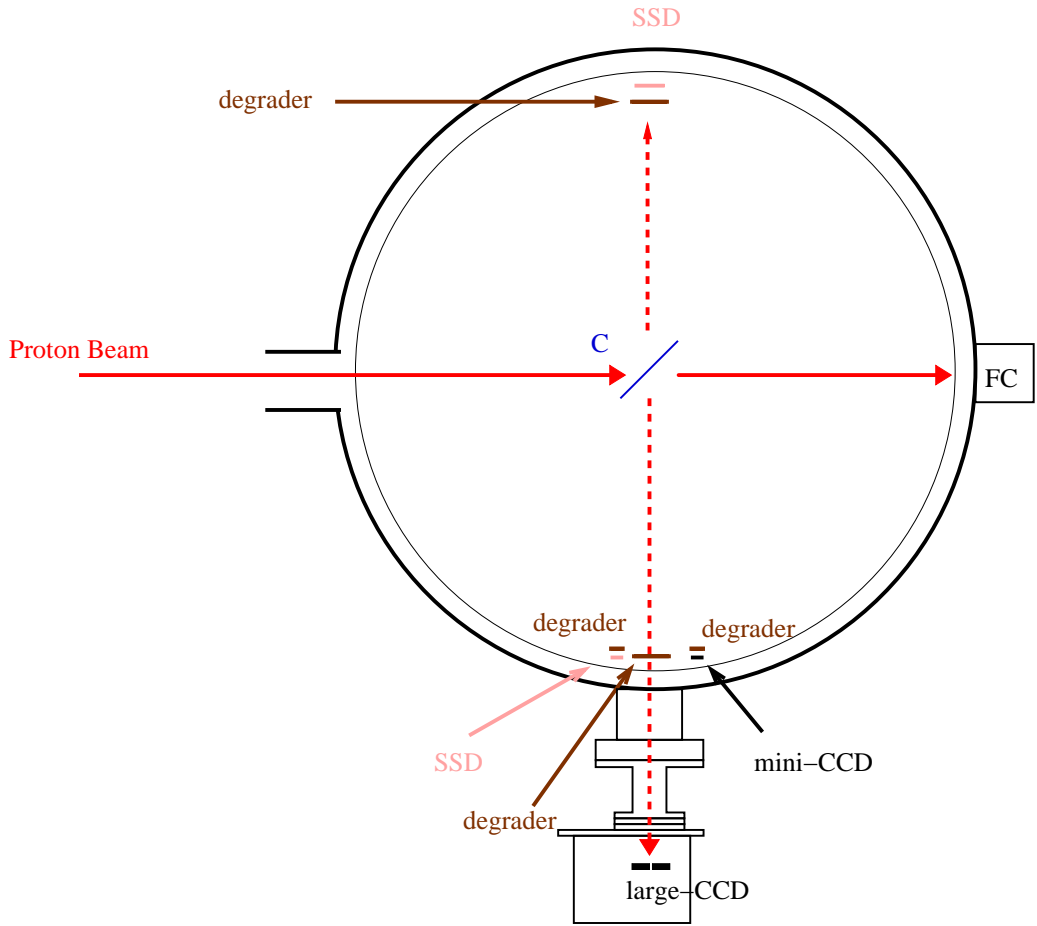}
 \caption[]{Top view of experimental setup in the scattering chamber of
 1~m in diameter. }
 \label{fig:chamber}
\end{figure}

Figure~\ref{fig:chamber} shows a top view of our experimental setup in a
scattering chamber of 1~m in diameter. The direct proton beam from the
accelerator was scattered at the center of the chamber. The large-CCD,
which was installed in a camera body attached outside of the chamber via
flange, was located at a right angle to the axis of the proton beam. The
proton beam intensity was measured by a Faraday cup, and the energy
spectra of the scattered protons were measured by silicon solid-state
detectors.
 
A thin carbon foil of 15~mg~cm$^{-2}$ was used as a scattering
target. The first excited energy level of $^{12}$C nucleus is about
4.4~MeV that is large enough to remove inelastic scattered protons using
a thin aluminum degrader shown in Figure~\ref{fig:chamber}.  The proton
beam energy was 10.5~MeV, and protons incident on the CCDs through
degraders were mono-energetic with a peak energy of 6.7~MeV and a full
width at half maximum of 0.9~MeV. The protons incident on the CCDs had a
range of about 360~$\mu$m in silicon and easily penetrated our device
with a thickness of 200~$\mu$m. Therefore, energy deposition was
relatively uniform along the depth direction with
12--16~keV~$\mu$m$^{-1}$. The total deposited energy in our device of 
single proton was about 2.7~MeV.

The large-CCD was operated in the camera body with a temperature of -110$^{\circ}$C
and the whole imaging area was irradiated with an $^{55}$Fe source. The back bias
voltage applied was 35~V. Degradation of the large-CCD was monitored alternating
proton irradiation and $^{55}$Fe data acquisition. On the other hand, the mini-CCD
was placed without a camera system inside the chamber hence its gradual change was
not monitored. Instead, the mini-CCD provided a higher dose data because of its
closeness to the scattering target than the large-CCD. The $^{55}$Fe data of the
mini-CCD were taken before and after this experiment in our laboratory.

\section{Dose rate in the orbit of the Astro-H satellite and exposed dose in this experiment } 

In order to determine whether or not our device has radiation hardness enough for
space use, it is necessary to estimate the dose rate in space and convert the
exposed dose in this experiment to equivalent time in space. For the calculation of
the dose rate, we referred to the day-averaged cosmic-ray flux model in the low
earth orbit at an inclination angle of $\sim$30$^{\circ}$\cite{2010SPIE.7732E.105M},
in which the Astro-H will fly. The cosmic-ray model shows that the
geomagnetically-trapped proton in the South Atlantic Anomaly (SAA) is by far the
largest population among high-energy particle and radiation components which can
penetrate a surrounding camera body and reach the CCD inside. At least more than
90\% of whole high-energy charged particles exposed to a satellite come from the SAA
on a day-average basis (T.\ Mizuno, private communication). We thus used the SAA
proton component alone from the cosmic-ray model.

\begin{figure}[ht]
 \centering \includegraphics[width=1.0\linewidth]{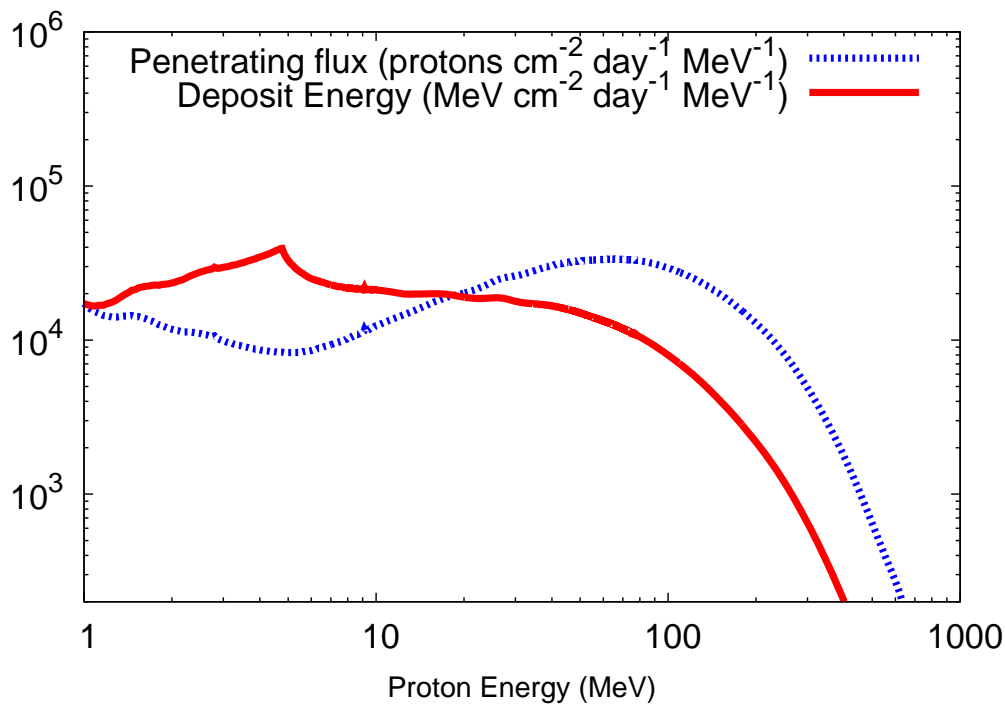}
 \caption[]{Number flux spectrum of protons that penetrate the camera body and
 reach the CCD (blue) and energy spectrum deposited from protons at the CCD (red).}
 \label{fig:spectra}
\end{figure}

Figure~\ref{fig:spectra} shows a number flux spectrum of protons that
penetrate the camera body and reach the CCD. This spectrum was obtained
from the SAA proton flux spectrum calculating their energy loss at the
passage of the camera body. We approximated the camera body as an
aluminum 2~cm thickness shell in this
calculation. Figure~\ref{fig:spectra} also shows an energy spectrum
deposited from protons at the CCD. This spectrum was obtained multiplying
the incoming proton number flux to the CCD with their energy deposition
at the CCD that is a function of the proton energy. The incoming number
flux spectrum has a peak around $\sim$70~MeV, which is the threshold 
energy to penetrate 2~cm thick aluminum camera body. On the
other hand, the deposited energy spectrum shows that lower energy proton
population contributes more to the total dose at the CCD. Integrating the
deposited energy spectrum, we obtained a dose rate of
2.2~$\times$~10$^{6}$~MeV~cm$^{-2}$~day$^{-1}$ or
260~rad~year$^{-1}$. Uncertainty of this value was estimated to be at
most a factor of 2 that mainly comes from the SAA proton flux spectrum
(T.\ Mizuno, private communication).

The total proton fluence exposed to the large-CCD and the mini-CCD were
about 0.9 and 3.7~$\times$~10$^{9}$~cm$^{-2}$, respectively. These
values correspond to about 3~and~13~year in orbit, which covers the
Astro-H satellite's mission life time of 5~years. The gradual
degradation of the large-CCD was monitored at equivalent times in orbit
of 2~days, 2~weeks, 2~months, 6~months, 1~year, 2~years, and 3~years.

\section{Result}

\subsection{Charge transfer inefficiency as a function of proton fluence}

\begin{figure}[h]
 \centering
 \centering \includegraphics[width=1.0\linewidth]{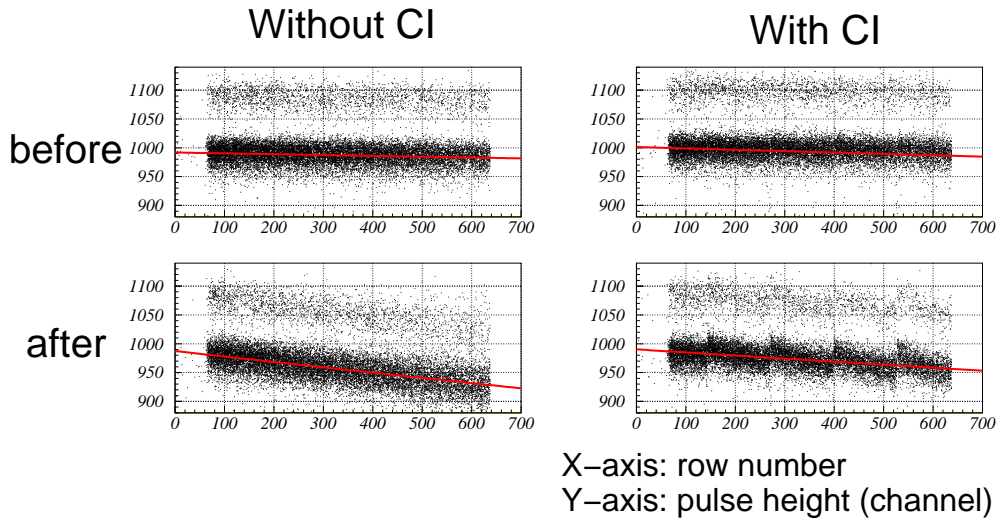}
 \caption[]{Stacking plots (black dots) with the best fit CTI function
 (red). Top and bottom figures show those before (no fluence) and after proton
 irradiation (0.9~$\times$~10$^{9}$~cm$^{-2}$), while left and right figures show
 those without or with CI.} 
 \label{fig:stacking}
\end{figure}

Figure~\ref{fig:stacking} shows ``stacking plots'' made from $^{55}$Fe
irradiation data, in which pulse heights of the X-ray hit pixels are plotted as
a function of row number that is namely half of the number of transfers in our
case because of on-chip 2$\times$2 binning. Each stacking plot shows dense and
faint bars which come from Mn~K$\alpha$ and Mn~K$\beta$ data, respectively. In
the analysis, we used selected events in which the charge is in a single pixel,
not be split into neighboring pixels. 

Before the experiment (no fluence), the pulse height decline is barely seen along
the row number. On the other hand, after the experiment (the proton fluence exposed
was 0.9~$\times$~10$^{9}$~cm$^{-2}$), the pulse height clearly decreases as a
function of the row number due to an increase of the CTI. The scatter of the pulse
height at a given row number also becomes larger, which is more evident in
Mn~K$\beta$ data in this figure. Comparing the bottom two plots, it is clear that
applying the CI technique reduces radiation damage effects of this type: both the
pulse height decline and scattering are mitigated. A ``saw-tooth shape'' seen in the
bottom-right plot is the characteristic of applying the
CI\cite{2009PASJ...61S...9U}. The degree of the mitigation is locally maximum just
after the charge-injection row and decreases as apart from it. The regularly-spaced
charge-injection rows thus make such a periodic pattern in the stacking plot.

The CTI value can be determined by fitting the stacking plots with the following
function,
\begin{align}
 Q &= Q_{0} \times (1- CTI)^{2Y},
\end{align}
where $Q_{0}$ is the original charge produced by X-ray from $^{55}$Fe, $Q$ is the
observed charge after transfer, and $Y$ is the row number of the pixel hit by the
X-ray. Red lines in Figure~\ref{fig:stacking} show the best fit curves only using
the Mn~K$\alpha$ data. The CTI value represents the slope of the stacking plot.

\begin{figure}[ht]
 \centering \includegraphics[width=0.95\linewidth]{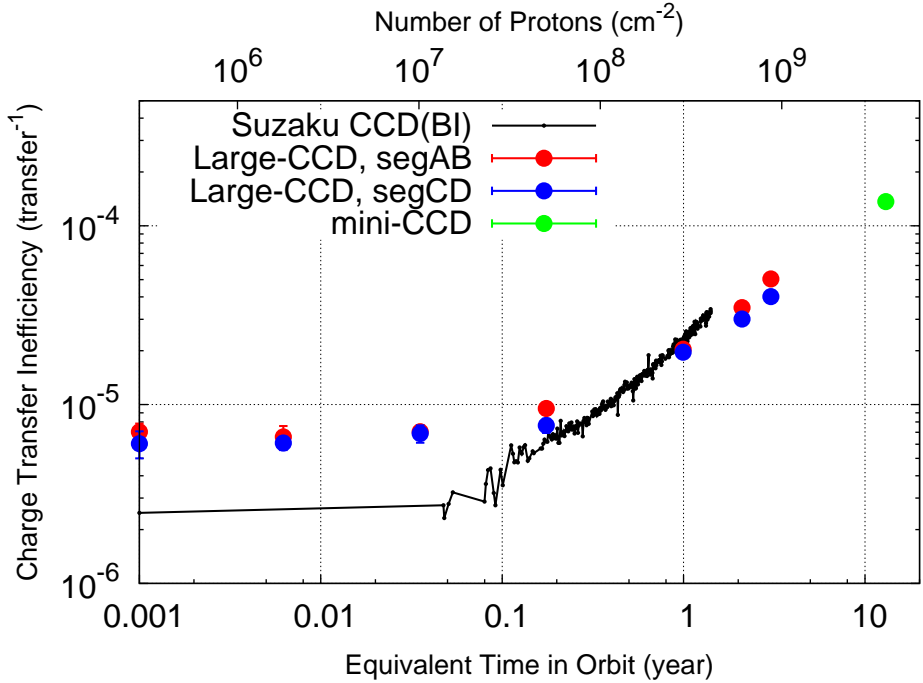}
 \caption[]{Charge transfer inefficiency measured without applying the CI
 technique. The data are plotted as a function of proton fluence (top label) or
 equivalent time in orbit (bottom label). Red and blue circles indicate the
 large-CCD data but from two different segments while green circles indicate the
 mini-CCD data. Black dots show Suzaku BI CCD data. Note that the data point
 equivalent to 6~month in orbit is lacking because of operation error.} \label{fig:CIoff}
\end{figure}

Figure~\ref{fig:CIoff} shows the CTI measured without applying the CI technique. The
data are plotted as a function of proton fluence (top label) or equivalent time in
orbit (bottom label). The colored circles indicate the data taken from our
experiment. There are two segments (we call AB and CD) in the large-CCD as we used
two readout nodes. Systematic uncertainties dominate and we estimated them as
follows. We first divided the data into 10 subsets, derived the CTI value for each
subset, and used a scatter of the values as a systematic uncertainty of the data,
although they were still smaller than the radius of the circles in the figure. The
black dots show the Suzaku BI CCD data actually taken in space. The Suzaku CCD has
two $^{55}$Fe calibration sources irradiating top corners of the imaging area (far
side from readout nodes)\cite{2007PASJ...59S..23K}. The calibration data was
available on a daily basis and summarized on-line\footnote{see
http://space.mit.edu/XIS/monitor/ccdperf/}. We reproduced the CTI values based on
the on-line data. Data points at 0.001~year equivalent in orbit mean those before
experiment or before launch.

In the case of our P-channel CCD, the CTI was measured to be 7.0~$\times$~10$^{-6}$
before experiment and more or less constant until about 1~month or so. It then went
up to an order of 10$^{-5}$ after 4.5 months, and reached to 5.0~$\times$~10$^{-5}$
at 3~year. The mini-CCD data point suggests that the CTI would rise with a similar
slope up to about 13~year. This time evolution is similar to that of the Suzaku CCD
data, especially in terms of the slope after a few months. The apparent difference
between their slopes is likely due to the possible uncertainty of factor of 2 in the
conversion of proton fluence to equivalent time in orbit.

\begin{figure}[ht]
 \centering
 \includegraphics[width=0.95\linewidth]{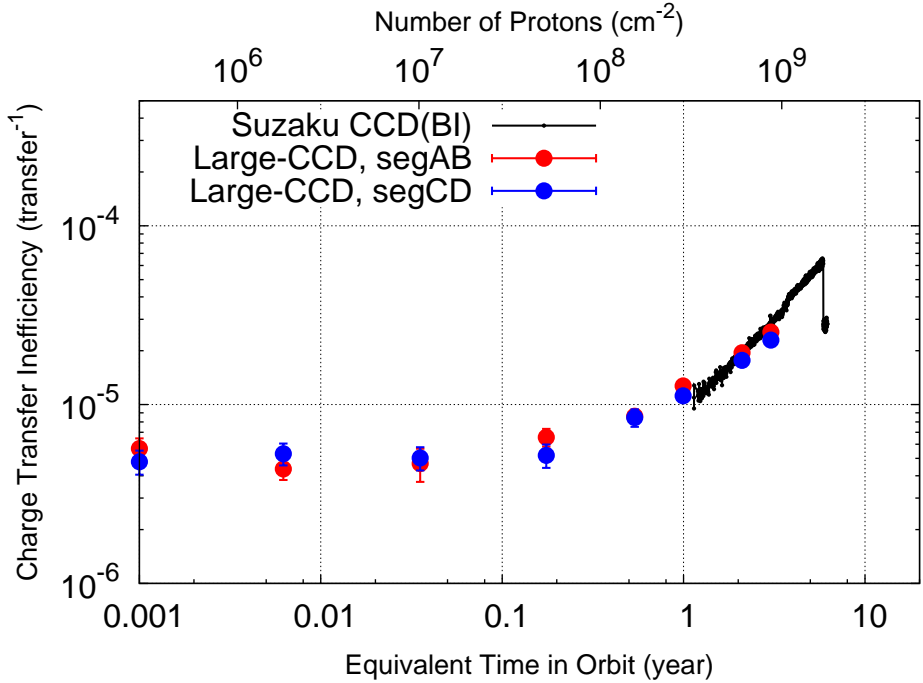}
 \caption[]{Charge transfer inefficiency measured with applying the CI
 technique.  The data are plotted as a function of proton fluence (top
 label) or equivalent time in orbit (bottom label). Red and blue circles
 indicate the large-CCD data but from two different segments. Black dots
 show Suzaku BI CCD data. A discontinuous decrease of the CTI in the
 Suzaku data at 6~years after the launch is due to an increase of the
 amount of injected charge from 2~keV equivalent to 6~keV equivalent.}
 \label{fig:CIon}
\end{figure}

Figure~\ref{fig:CIon} shows the CTI measured with applying the CI
technique. The definition of the marks and labels are the same as that
of Figure~\ref{fig:CIoff}. The mini-CCD data with the CI was not
available. The Suzaku CCD had been operated without the CI until
1.2~year after the launch. After some verification
tests\cite{2007SPIE.6686E..23B}, the CI was incorporated in the standard
observation mode. Thus, the calibration data with the CI were also only
available since about 1~year after the launch. The amount of injected
charge was increased from 2~keV equivalent to 6~keV equivalent about
6~years after the launch, which reduced the CTI as shown in
Figure~\ref{fig:CIon}.

Applying the CI technique reduces not only the CTI values themselves but
also its growth rate: the reduction factors are 1.5, 1.8, and 2.0 in the
last three data. This suggests that the CI works more efficiently when
the device is more damaged at least in our case. The slope is again
similar to that of the Suzaku CCD data.  Our P-channel data appear to go
along with the 6~keV equivalent Suzaku CCD data rather than that of the
2~keV equivalent, although it may not be significant considering the
uncertainty.

In the both cases with or without the CI, the degradation of the CCD
performance in terms of the CTI is shown to be comparable with that of
the Suzaku CCD. Applying the same ground-base CTI correction as those
for the Suzaku CCD data to our P-channel CCD data, we can expect to
provide the data with the similar quality as the Suzaku data. 

\subsection{Charge transfer inefficiency as a function of temperature}
\label{sec:tdependency}

\begin{figure}[ht]
 \centering
 \includegraphics[width=1.0\linewidth]{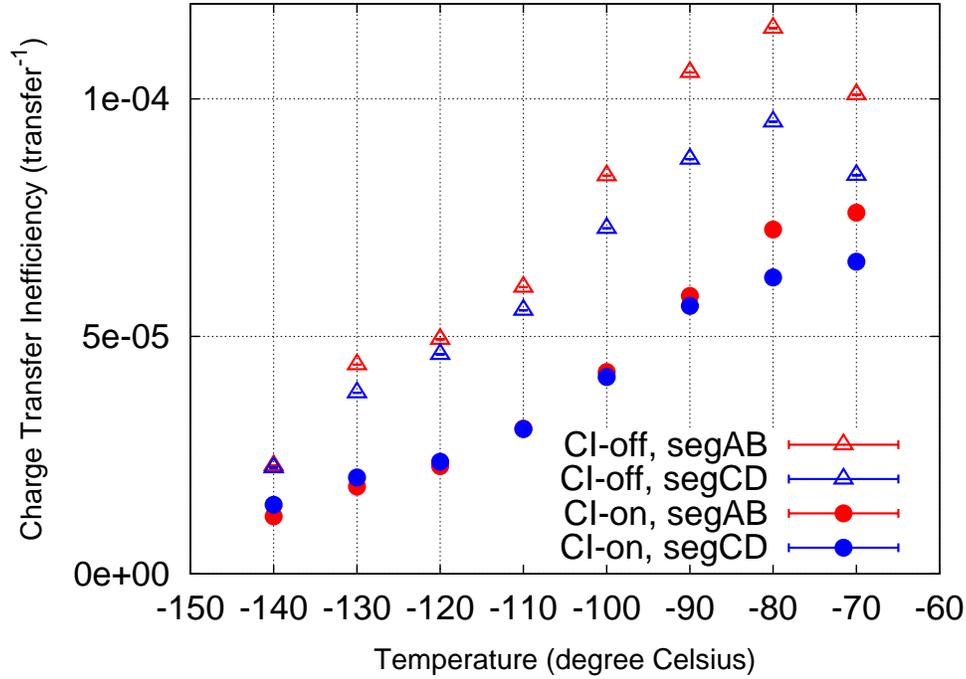}
 \caption[]{Charge transfer inefficiency measured as a function of the
 temperature of the CCD. Triangles and circles indicate the data without and
 with the CI, respectively. The different color indicate different segment data.}
 \label{fig:tdependency}
\end{figure}

Figure~\ref{fig:tdependency} shows the CTI measured as a function of the temperature
of the CCD. These data were taken in our laboratory after the proton radiation
damage experiment. The values are slightly different from those shown in previous
figures even at the same temperature (-110~$^{\circ}$C) probably because of
partially used different electronics. Within the temperature range we tested, it is
found that the CTI is smaller at lower temperature. The turn over of the data
without the CI at -80$^{\circ}$C is not real. At temperature higher than
-90$^{\circ}$C, a charge-trailing became obvious likely due to an increase of the
number of traps with a short detrapping time scale of a few CCD clock
cycles\cite{2007PASJ...59S..23K}. In such a situation, using a single pixel data
introduces a selective bias and the CTI would be falsely evaluated to be
smaller. The difference between the two segments is also smaller at the lower
temperature. The effect of applying the CI technique reducing the CTI by about a
factor of 2 is almost the same regardless of the temperature.

\section{Discussion}

We performed a proton radiation damage experiment on our newly developed
P-channel CCD and confirmed that its radiation hardness is comparable with that
of the Suzaku N-channel CCD actually working in space. Although this result
verifies the validity of space use of our P-channel CCD, it might contradict
with the previous report that P-channel CCDs are radiation harder than N-channel
CCDs.

\begin{figure}[ht]
 \centering \includegraphics[width=1.0\linewidth]{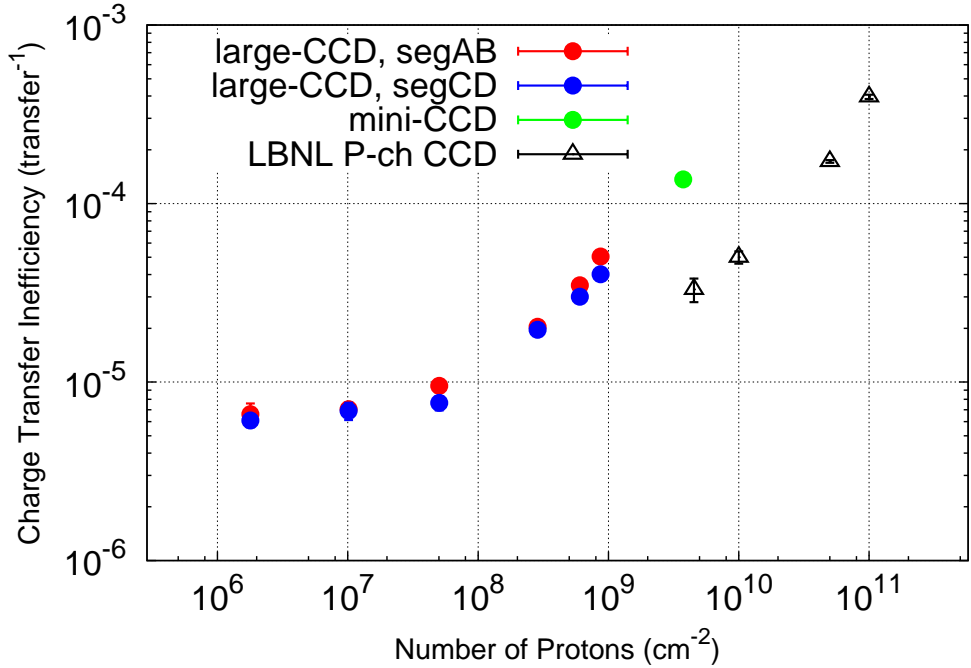}
 \caption[]{A comparison of the radiation hardness of our P-channel CCD without
 applying the CI (colored circles) and the LBNL P-channel CCD without the notch
 structure (black triangles). The data of the LBNL P-channel CCD are taken from
 Bebek et al.\ (2002)\cite{2002ITNS...49..1221N}.}  \label{fig:bebek}
\end{figure}

Figure~\ref{fig:bebek} shows a comparison between our results and those reported
from another team using a P-channel CCD made by the Lawrence Berkeley National
Laboratory (LBNL) \cite{2002ITNS...49..1221N}. There were two types of the LBNL
CCDs whose data were published: one had the ``notch'' structure and the other
did not. The notch structure, a narrow implant in the CCD channel confining a
charge packet to a fraction of the pixel volume in an additional potential
well, has been known to reduce the
CTI\cite{2004ITNS...51..2288N,2002ITNS...49..1221N}. In order to make a direct
comparison, we plot our P-channel CCD data without applying the CI and the LBNL
P-channel CCD data without the notch structure.  A simple comparison along the
number of protons irradiated indicates that the LBNL CCD may have about an order
of magnitude higher radiation tolerance compared to our CCD. This difference can
not be resolved even considering the differences in the incident proton energies
and the CCD temperatures between the two experiments. In the LBNL CCD
experiment, the incident proton energy and the CCD temperature were 12~MeV and
-145$^{\circ}$C, respectively. The energy deposit of the 12~MeV protons is about
a factor of 2 lower than that of 6.7~MeV protons used in our experiment. From
figure~\ref{fig:tdependency}, lowering the CCD temperature from -110$^{\circ}$C
to -145$^{\circ}$C would reduce the CTI by about a factor of 2. Thus, about a
factor of 4 difference can be explained by the difference in the experimental
setup. However, significant difference in the radiation hardness still remains
between the two P-channel CCDs. The different manufacturing process might be a
reason but we do not have a clear answer at this moment.

The LBNL CCD experiment was performed while the devices were unpowered and at room
temperature. On the other hand, the large-CCD was powered and cooled during our
proton irradiation experiment. However, this difference does not appear to affect
the degree of the damage at least in terms of the CTI, considering the relatively
smooth connection between the large-CCD and the mini-CCD data. The same conclusion
was deduced for N-channel CCDs in a similar proton radiation
experiment\cite{2002JaJAP..41.7542M}.

The temperature dependency measurement described in section~\ref{sec:tdependency}
was performed in our laboratory about two months after the proton radiation damage
experiment. Meanwhile the large-CCD was kept at room temperature. A CTI measurement
at the same CCD temperature showed about a 40\% worse value compared to the last
value obtained at the accelerator laboratory. As we previously noted, this apparent
degradation was most likely due to a partial difference of electronics used. This
fact then suggests that an annealing at room temperature did not help to restore the
radiation damage for our device at least at this damage level.

We confirmed that applying the CI technique and lowering the CCD temperate are
both efficient methods to mitigate the radiation damage.  Although we
tentatively spaced the CI rows 128 row apart in this experiment, the Suzaku CCD
camera injects charges at every 54 row. We can expect further reduction of the
CTI by narrowing the spacing of the CI rows. Since too many CI rows also reduce
an effective imaging area, we need to find out an optimized spacing in future
experiments.

The temperature dependence of the CTI of the damaged CCDs varies device to
device\cite{2004SPIE.5488..264S,2006SPIE.6276E..48G}. Therefore, the
understating of the temperature dependence allows us a flexible operation in
space where the power available is quite limited. For example, it is a possible
option for us to operate the CCD with -100$^{\circ}$C in the initial phase and
-120$^{\circ}$C in the later phase of the mission based on this result.

\section*{Acknowledgment}

We would like to thank all the SXI team members. We also appreciate Dr.\ T.\
Mizuno who kindly provided the cosmic-ray flux model used in this report, Dr.\
M.\ Kokubun for useful comments, and the Suzaku XIS team at MIT for making the
$^{55}$Fe monitoring data available on-line. This work was supported by JSPS
KAKENHI Grand Number 23000004.

\end{document}